\documentclass[aip,superscriptaddress,apl,reprint]{revtex4-1}
\usepackage{graphicx}
\usepackage{amssymb}
\begin{document}

\title{Single-layer metallicity and interface magnetism of epitaxial graphene on SiC(000$\bar{1}$)}

\author{I. Deretzis}
\email{ioannis.deretzis@imm.cnr.it}
\affiliation{CNR-IMM, Z.I. Ottava Strada 5, 95121 Catania, Italy}

\author{A. La Magna}
\affiliation{CNR-IMM, Z.I. Ottava Strada 5, 95121 Catania, Italy}
\date{\today}

\begin{abstract}
We perform density functional theory calculations for the determination of the structural and electronic properties of epitaxial graphene on 4H-SiC(000$\bar{1}$). Using commensurate supercells that minimize non-physical stresses we show that, in contrast with Si-face epitaxial films, the first graphene layer that forms on the C-face of SiC is purely metallic with its $\pi$-bands partially preserved. Typical free-standing characteristics are fully recovered with a second graphene layer. We moreover discuss on the magnetic properties of the interface and the absence of Fermi-level pinning effects that could allow for a plausible device operation starting from the off-state.

\end{abstract}

\maketitle


Epitaxial graphene has emerged as a highly attractive system for device integration by providing a combination of characteristics that constitute a significant advantage with respect to competitive graphene production technology: wafer size scales\cite{2009NatMa...8..203E} and direct growth on semi-insulating substrates. The process is based on the sublimation of Si atoms starting from SiC substrates in ultra high vacuum or ambient pressure furnaces\cite{2009NanoL...9.2605J}. Accurate control of the growth parameters can allow for the reformulation of graphene films from the remaining C surface atoms. The structural, electronic and transport properties of epitaxial graphene strongly depend on the the polarization of the SiC surface: Si-face epitaxial graphene (i.e. graphene on SiC(0001)) is characterized by the formation of a first carbon-rich interface layer with a $6\sqrt3\times6\sqrt3R30^{\circ}$ surface reconstruction, over which Bernel stacked graphene layers grow. Interface interaction imposes significant $n$-type doping while scatterers reduce mobility values with respect to the $SiO_2$-deposited case\cite{2009NatMa...8..203E}. Typical graphene characteristics (e.g. the half-integer quantum hall effect) are recovered by the application of a gate voltage that lowers the Fermi level around the Dirac point\cite{2010PhRvB..81s5434J}. C-face epitaxial graphene (i.e. graphene on SiC(000$\bar{1}$)) is subject to a less stringent rotational ordering with respect to the substrate\cite{2010arXiv1001.3869S,2009JPCM...21m4016S} whereas the presence of an interface buffer layer, although predicted by theoretical calculations\cite{2007PhRvL..99g6802M,2007PhRvL..99l6805V}, is still a matter of debate\cite{camara:093107,2007PhRvB..75u4109H,2009PhRvB..80w5429H,2009JPCM...21m4016S}. A complex rotational symmetry of subsequent graphene layers other than the AB stacking sequence is present \cite{2010arXiv1001.3869S} that allows for the manifestation of single-layer properties even in the case of a multilayer structure\cite{2009PhRvL.103v6803S}. C-face monolayers show higher mobilities with respect to the (0001) case and the typical half-integer quantum hall effect at low temperatures\cite{camara:093107,2009ApPhL..95v3108W}. Doping is also present here, however controversy holds over the type and the concentration of carriers\cite{2010PhRvL.104m6802S,camara:093107,2009ApPhL..95v3108W}.

In this article we focus on C-face epitaxial graphene by means of density functional theory calculations. Starting from a basis of lattice commensuration for graphene and the SiC substrate, we show that results can significantly differ with respect to non-commensurate models\cite{2007PhRvL..99g6802M,2007PhRvL..99l6805V}. Contrary to the Si-face\cite{2009ApPhL..95f3111D}, we find that the first graphene layer preserves a purely metallic character with a complete absence of a bandgap, while partially maintaining its $\pi$-bands. However, interaction with the substrate perturbs the typical Dirac cone that only appears with the addition of a second graphene layer. We moreover show the presence interface ferromagnetism due to strong exchange interactions and an absence of the Fermi-level pinning effects previously reported for (0001) films\cite{2009ApPhL..95f3111D,2009PhRvB..80x1406S,ouerghi:161905}.


\begin{figure}
	\centering
		\includegraphics[width=\columnwidth]{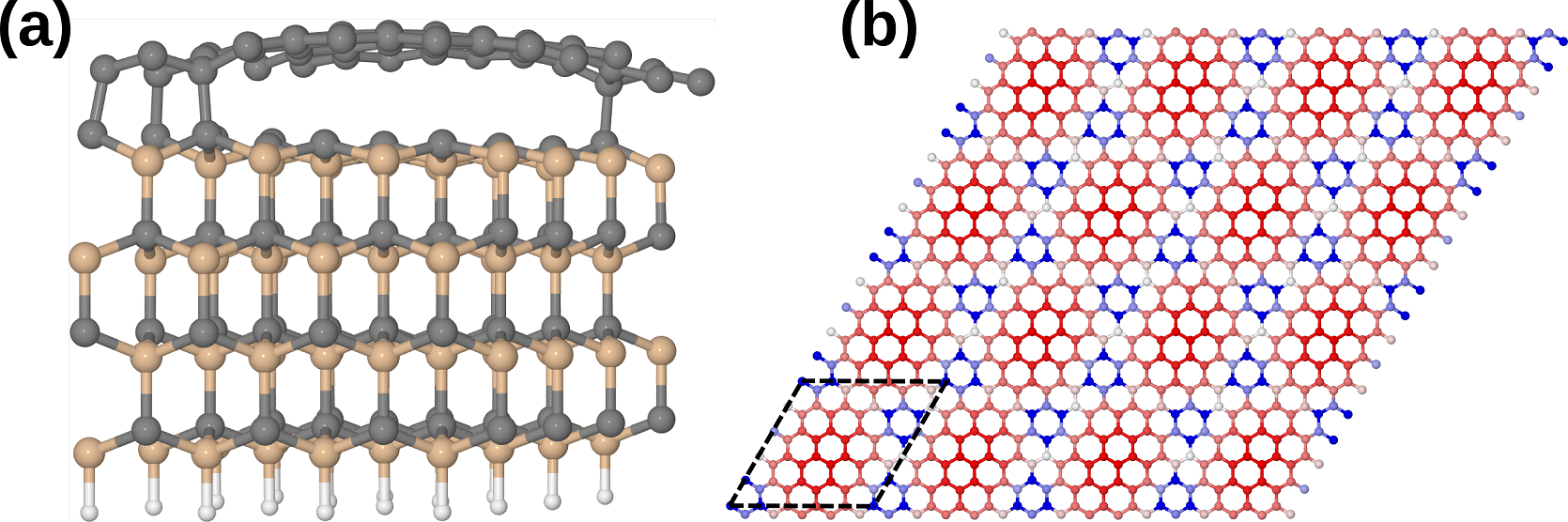}
	\caption{(a) Side view of a ($5 \times 5$) graphene monolayer relaxed on a ($4 \times 4$) 4H-SiC(000$\bar{1}$) substrate. (b) Color map view of the first epitaxial graphene layer showing the bonding characteristics with the substrate, where gradual red to blue coloring indicates $sp^2$ to $sp^3$ bonding. Dashed lines show the periodically reproduced unit cell.}
	\label{fig:geometry}
\end{figure}

We perform \textit{ab initio} calculations based on the Density Functional Theory (DFT) within the Local Spin Density Approximation (LSDA) as implemented in the SIESTA computational code\cite{2002JPCM...14.2745S}. The 4H-SiC epitaxial graphene structures comprise of four bilayers of a ($4 \times 4$) SiC substrate (passivated with H at the bottom) over which single and double layers of graphene grow, forming a ($5 \times 5$) supercell that satisfies lattice commensuration and minimizes non-physical stresses (Calculated lattice constants: $a_{SiC} = 3.076$\AA, $a_{gr} = 2.46$\AA). Contrary to $R30^{\circ}$ reconstructions, here there exists no rotational angle for the graphene supercell with respect to the SiC  substrate. Such a configuration is often observed in experiments for (000$\bar{1}$) samples\cite{2010arXiv1001.3869S}. A basis set of double-$\zeta$ valence (plus polarization) orbitals has been used for C (Si) and H, along with Troulier-Martins pseudopotentials\cite{1991PhRvB..43.1993T} for the modeling of ionic cores. Both basis set sensitivity and pseudopotentials have been tested to accurately reproduce the bandstructure of hexagonal SiC polytypes and graphene. Sampling of the Brilouin zone takes place by a $2 \times 2 \times 1$ Monhorst-Pack grid, whereas bandstructure is plotted for 130 nonequivalent $k$-points along the closed $\Gamma \rightarrow M \rightarrow K \rightarrow \Gamma$ path. A mesh cutoff energy of 500 Ry has been imposed for real-space integration, while structures have been relaxed until forces were less than 0.04 eV/\AA{}. 

\begin{figure}
	\centering
		\includegraphics[width=\columnwidth]{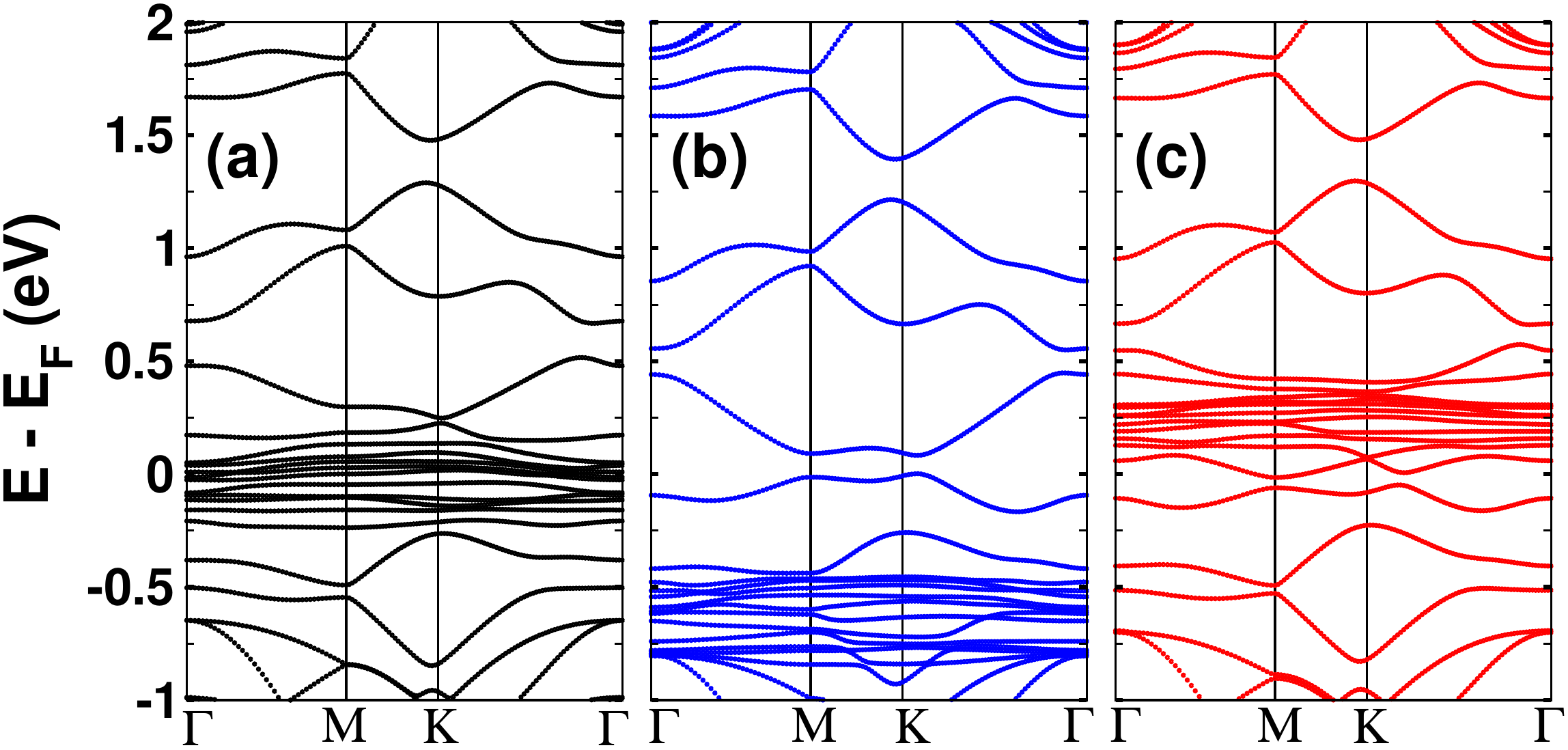}
	\caption{Bandstructure of a graphene monolayer on a 4H-SiC(000$\bar{1}$) substrate: (a) spin-restricted configuration, (b) majority-spin configuration and (c) minority-spin configuration.}
	\label{fig:bs1}
\end{figure}

At the end of the relaxation process, the first graphene layer presents a corrugation with thickness $ t \approx 1 $ \AA{} (Fig. \ref{fig:geometry}(a)) and has structural characteristics that present some fundamental differences with respect to the buffer layer of the non-commensurate $\sqrt3 \times \sqrt3 R30^{\circ}$ reconstruction\cite{2007PhRvL..99g6802M,2007PhRvL..99l6805V}. A periodic reproduction of the unit cell (Fig. \ref{fig:geometry}(b)) shows the presence of small islands covering almost the four fifths of the epilayer surface that strongly maintain $sp^2$-bonding characteristics and represent the areas of the honeycomb lattice that are more distant from the substrate ($ d_{max} \approx 3 $ \AA{}). Below these islands the substrate relaxes in an ideal surface reconstruction of the C-face\cite{2010JPCM...22z5003W} where also substrate carbon atoms strongly $sp^2$ hybridize. Substrate reconstruction is the origin of the energetic stability of this structure: we have calculated that the ground state energy proposed by our model is significantly lower than the one of the $\sqrt3 \times \sqrt3 R30^{\circ}$ case. The $sp^2$ islands of the epilayer are terminated by carbon atoms that strongly bond with the substrate and loose the $sp^2$ hybridization for the $sp^3$ one. From a modeling point of view, the increase of the lattice periodicity for this supercell with respect to the $\sqrt3 \times \sqrt3 R30^{\circ}$ one is in favor of $sp^2$ bonding. Indeed, the biggest part of carbon atoms that form the graphene epilayer maintain their $\pi$ orbitals and only few atoms covalently bond with the substrate. It should be moreover noted that in this geometrical configuration stresses are minimized, since the $sp^2$ bonds of the graphene layer have a mean distance of $C-C_{sp^2} \approx 1.43 $ \AA{} whereas interface $sp^3$ ones relax at $C-C_{sp^3} \approx 1.68 $ \AA{}, which slightly exceeds that of pure diamond. Moreover, interesting structural comparisons between the present calculations and the scanning tunneling microscopy images of Refs. \citenum{2009PhRvB..80w5429H,2009JPCM...21m4016S} for monolayer graphene grown on SiC (000$\bar{1}$)($3 \times 3$) can be made (e.g. see figure \ref{fig:geometry}(b) of the present manuscript and figure 2(d) of Ref. \citenum{2009PhRvB..80w5429H}).

\begin{figure}
	\centering
		\includegraphics[width=\columnwidth]{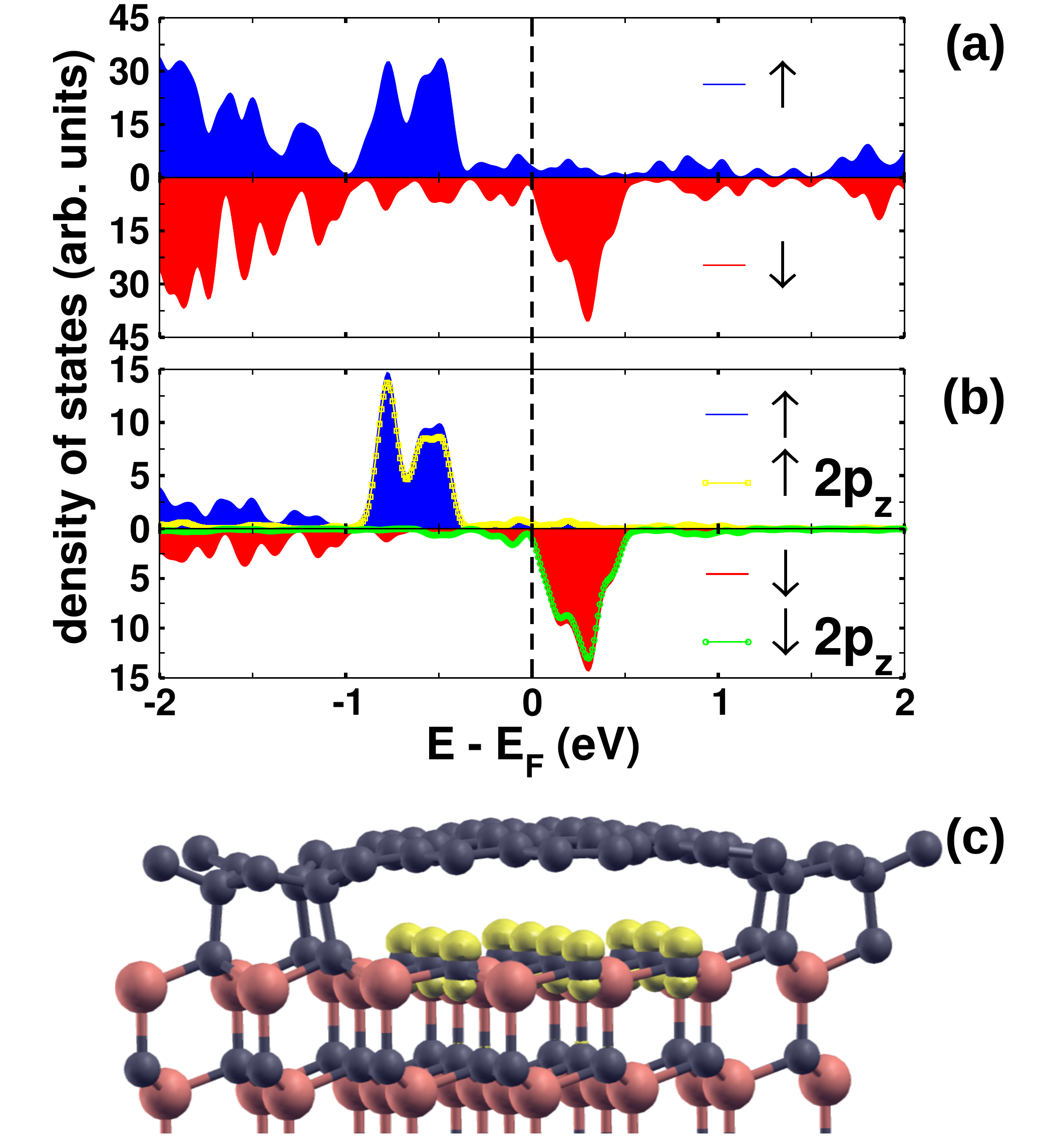}
	\caption{(a) Total density of states of the majority-spin (upper) and the minority-spin (lower) for the graphene/4H-SiC(000$\bar{1}$) system. (b) Projected density of states for the majority-spin (upper) and the minority-spin (lower) on the carbon atoms of the surface SiC bilayer below the graphene layer (lines). The contribution of the $2p_z$ orbitals is also shown (lines+symbols). (c) Local magnetization $(\rho_{\uparrow} - \rho_{\downarrow})$ for the previous system, where yellow isosurfaces indicate an excess of $\uparrow$ electrons.}
	\label{fig:pdos}
\end{figure}

Fig. \ref{fig:bs1} shows the electronic bandstructure of the first graphene layer on 4H-SiC(000$\bar{1}$). The key aspect of this interaction is that the monolayer preserves a purely metallic character with its $\pi$ bands prevailing within the SiC-substrate bandgap, in consistency with electrical measurements on C-face monolayers\cite{2009PhRvB..80w5429H}. This feature is in a clear contrast with the carbon-rich layer grown on the Si-face\cite{2007PhRvL..99l6805V}. However, the interaction with the substrate, notwithstanding weak, perturbs the Dirac cone and the $\pi$ bands do not preserve the typical linearity of free-standing graphene near the Fermi level. A careful view of the spin-polarized bandstructure (Fig. \ref{fig:bs1}(b-c)) that corresponds to the most stable electronic configuration shows some important differences with respect to the spin-unpolarized one (Fig. \ref{fig:bs1}(a)). The presence of quasi-flat bands near the Fermi level of this system derives from SiC surface and is localized on C atoms that are not bound to the graphene epilayer. From the spin-unpolarized bandstructure one can deduce that these states pin the Fermi level at the interface as similarly calculated for Si-face grown graphene\cite{2009ApPhL..95f3111D} (where this phenomenon is argued to be one of the main factors for the strong $n$-type doping of that system\cite{2009PhRvB..80x1406S,ouerghi:161905}). However the spin-polarized bandstructure shows a different picture: here the surface states are half-filled and strong exchange interactions make them split above and below the Fermi level by $\sim 0.7-0.8 eV$, leaving the graphene/4H-SiC(000$\bar{1}$) system unpinned (in agreement with previous calculations of Ref. \citenum{2007PhRvL..99g6802M}). 

The presence of significant exchange interactions makes necessary a more careful analysis of magnetism issues in the interface of these systems. Fig. \ref{fig:pdos} shows spin-up ($\uparrow$) and spin-down ($\downarrow$) density of states (DOS) and electronic density $\rho$ configurations for the  graphene/4H-SiC(000$\bar{1}$) system. The total density of states (Fig. \ref{fig:pdos}(a)) reveals the presence of a big concentration of spin-up states from $0.3 eV$ to $1 eV$ below the Fermi level, and similarly for spin-down states from $0 eV$ to $0.5 eV$ above the Fermi level. A careful analysis of the projected density of states on the carbon atoms of the first SiC bilayer below the graphene layer (Fig. \ref{fig:pdos}(b)) shows that a major component of these two peaks derives from the those carbon atoms that are not covalently bonded with the graphene overlayer. Moreover, the contributions of the $2p_z$ orbitals are almost exclusive in these peaks (the remaining DOS contributed by $2p_z$ orbitals that can be found in the system is localized at the graphene monolayer). The surface $\uparrow$ and $\downarrow$ band-splitting gives rise to local magnetization phenomena at the heterostructure's interface (Fig. \ref{fig:pdos}(c)) where a ferromagnetic order is present. C substrate atoms that are not covalently bound with the epilayer have magnetic moments that range within $0.6 - 0.67 \mu_B$, whereas those covalently bound are not magnetic. Magnetism holds only for the first bilayer of the substrate and rapidly decreases with distance: the third SiC bilayer and the graphene epilayers are not magnetic. Interface magnetism however influences the electronic bands of the graphene epilayer where a shift of $\sim 0.1 eV$ between $\uparrow$ and $\downarrow$ bands can be observed (Fig. \ref{fig:bs1}(b) and \ref{fig:bs1}(c)). 

\begin{figure}
	\centering
		\includegraphics[width=\columnwidth]{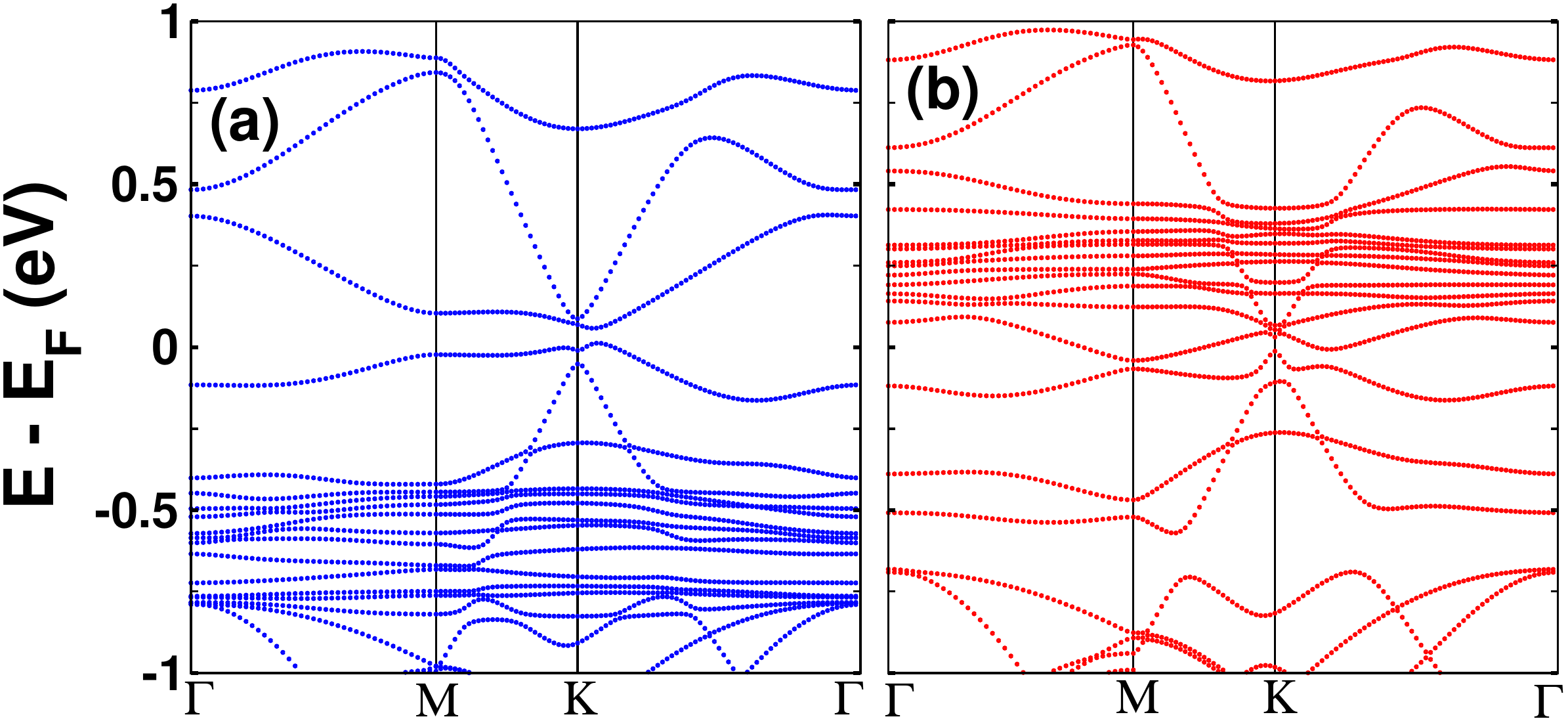}
	\caption{Bandstructure of two graphene layers on a 4H-SiC(000$\bar{1}$) substrate for (a) the majority-spin and (b) the minority-spin configuration.}
	\label{fig:bs2}
\end{figure}

As in the case of (0001) epitaxy, typical free-standing graphene characteristics and the presence of the Dirac cone are recovered with the second graphene layer (Fig. \ref{fig:bs2}). This layer interacts weakly with the first epilayer, maintaining a mean distance that is similar to that of Bernel-stacked graphite and shows a smaller corrugation with a layer thickness $t \sim 0.4$ \AA{}, in accordance with measurements\cite{2007PhRvB..75u4109H}. It should be again noted here that the Fermi level remains unpinned by the interface and is purely determined by the position of the Dirac cone. 

To conclude, in this study we have presented \textit{ab initio} electronic structure calculations for monolayer and bilayer epitaxial graphene systems grown on 4H-SiC(000$\bar{1}$). Contrary to epitaxial graphene on SiC(0001), the first graphene layer that grows on the C-face of SiC maintains a purely metallic character and an important presence of $\pi$ electrons along with a non-negligible coupling with the substrate. The particularity of its geometrical configuration consists in a corrugated surface where small $sp^2$-bonded islands are terminated by carbon atoms that covalently bind to the substrate. Below these islands also the surface carbon atoms $sp^2$-hybridize, while their $\pi$ bands are half-filled and present a ferromagnetic order due to the presence of strong electron exchange interactions. Typical graphene-like characteristics are recovered with the addition of a second graphene layer. According to this picture C-face epitaxial graphene should maintain some important advantages with respect to the Si-face one, like the absence of a Fermi-level pinning effect that could allow for a plausible device operation starting from an off-state. In this sense, doping effects measured in the laboratory\cite{2009ApPhL..95v3108W} should likely be attributed to environmental or local disorder factors like the presence of impurities, rather than considered as an intrinsic characteristic of the interface. Methodologically, this study argues that approaches that go beyond the $\sqrt3 \times \sqrt3 R30^{\circ}$ reconstruction\cite{2010PhRvB..82l1416P} are fundamental for the correct modeling of graphene on SiC(000$\bar{1}$).

The authors would like to acknowledge the European Science Foundation (ESF) under the EUROCORES Program EuroGRAPHENE CRP GRAPHIC-RF for partial financial support. Computations have been performed at the CINECA supercomputing facilities under project TRAGRAPH.

\end{document}